\documentclass[10pt]{article}
\usepackage[margin=1in]{geometry}
\usepackage[T1]{fontenc}
\usepackage[utf8]{inputenc}
\usepackage{graphicx}
\usepackage{booktabs}
\usepackage{multirow}
\usepackage{array}
\usepackage{amsmath,amssymb}
\usepackage{caption}
\usepackage{float}
\usepackage{longtable}
\usepackage{hyperref}
\usepackage{url}
\usepackage{booktabs}
\usepackage{siunitx}
\usepackage{makecell}
\usepackage[numbers,sort&compress]{natbib}
\usepackage{bm}
\title{PanGuide3D: Cohort-Robust Pancreas Tumor Segmentation via Probabilistic Pancreas Conditioning and a Transformer Bottleneck}

\author{%
Sunny Joy Ma$^{1}$, Xiang Ma$^{2,3,*}$\\[0.5em]
\small $^{1}$Trailridge School, Waukee, IA, 50263, USA\\
\small $^{2}$Department of Biochemistry, Grand View University, Des Moines, IA 50316, USA\\
\small $^{3}$Department of Computer Science, Iowa State University, Ames, IA 50011, USA\\
\small $^{*}$Correspondence: \href{mailto:xma@grandview.edu}{xma@grandview.edu}
}

\date{}

\begin{document}
\raggedright
\maketitle
\thispagestyle{plain}

\begin{abstract}
Pancreatic tumor segmentation in contrast-enhanced computed tomography (CT) is clinically important yet technically challenging: lesions are often small, heterogeneous, and easily confused with surrounding soft tissue, and models that perform well on one cohort frequently degrade under cohort shift. Our goal is to improve cross-cohort generalization while keeping the model architecture simple, efficient, and practical for 3D CT segmentation. We introduce PanGuide3D, a cohort-robust architecture with a shared 3D encoder, a pancreas decoder that predicts a probabilistic pancreas map, and a tumor decoder that is explicitly conditioned on this pancreas probability at multiple scales via differentiable soft gating. To capture long-range context under distribution shift, we further add a lightweight Transformer bottleneck in the U-Net bottleneck representation. We evaluate cohort transfer by training on the PanTS (Pancreatic Tumor Segmentation) cohort and testing both in-cohort (PanTS) and out-of-cohort on MSD (Medical Segmentation Decathlon) Task07 Pancreas, using matched preprocessing and training protocols across strong baselines. We collect voxel-level segmentation metrics, patient-level tumor detection, subgroup analyses by tumor size and anatomical location, volume-conditioned performance analyses, and calibration measurements to assess reliability. Across the evaluated models, PanGuide3D achieves the best overall tumor performance and shows improved cross-cohort generalization, particularly for small tumors and challenging anatomical locations, while reducing anatomically implausible false positives. These findings support probabilistic anatomical conditioning as a practical strategy for improving cross-cohort robustness in an end-to-end model and suggest potential utility for contouring support, treatment planning, and multi-institutional studies.
\end{abstract}

\section*{Introduction}
Pancreatic tumor segmentation in contrast-enhanced CT is a foundational step for downstream clinical workflows such as diagnosis support, treatment planning, and longitudinal response assessment\cite{zhang2024state, suneja2026systematic, arshad2026radiomics, lopez2025early, yao2020advances}. Yet it remains technically difficult even for modern 3D segmentation systems. Pancreatic lesions are often small relative to the organ, exhibit heterogeneous enhancement patterns, and can be visually confounded with nearby vessels and soft tissue~\cite{ghorpade2023review}. These properties create a severe small-target, low-signal regime with extreme class imbalance, where models must simultaneously avoid missing subtle tumors and suppress spurious detections that appear plausible in local texture but are anatomically implausible.

A key reason this problem persists is that real-world performance is governed not only by in-cohort accuracy but also by robustness to cohort shift. Differences in scanner vendors, reconstruction kernels, contrast timing, slice thickness, and patient populations can alter intensity statistics and lesion appearance, while annotation conventions and tumor prevalence can change the effective decision boundary. As a result, models that appear strong on a single benchmark often degrade sharply when evaluated on a new cohort, exhibiting reduced sensitivity on small tumors, inflated false positives, or unstable operating thresholds. Despite this practical reality, cross-cohort generalization is still under-emphasized relative to in-distribution benchmark performance, leaving a gap between reported results and deployable reliability.

Prior work has frequently addressed pancreas tumor segmentation by leveraging the pancreas as an anatomical prior. A common strategy is an organ-guided cascade: first segment or localize the pancreas, then restrict tumor prediction to a pancreas-centered region of interest or run a second model conditioned on the organ mask~\cite{qiu2024cascade,ghorpade2024twostage,mahmoudi2022pdac}. This approach reduces the search space and improves class balance, and it often suppresses false positives that occur outside plausible anatomy. In parallel, multi-task and multi-head U-Net variants have explored sharing representations between organ and lesion prediction, capitalizing on shared low-level features while allowing task-specific decoding~\cite{ghorpade2024twostage}. However, cascaded designs can be computationally expensive and brittle due to error propagation and hard ROI decisions, while generic multi-head designs may not explicitly encode organ uncertainty in a way that reliably constrains tumor predictions under distribution shift.

Motivated by these observations, we revisit organ guidance through a soft and end-to-end lens. Instead of treating pancreas prediction as a hard prerequisite that gates downstream processing, we model it as a probabilistic anatomical prior that continuously conditions tumor decoding. Concretely, a pancreas decoder produces a probability map that is injected into the tumor pathway at multiple scales, enabling the tumor decoder to emphasize pancreas-consistent features while still retaining the flexibility to recover from imperfect organ predictions. This probabilistic conditioning is differentiable, avoids brittle cropping decisions, and aligns with the intuition that what matters for tumor segmentation is not a binary pancreas mask but a calibrated notion of where pancreas likely is, especially when cohort shift changes contrast patterns and boundary clarity.

At the same time, robustness under cohort shift often requires more than local texture modeling. Convolutional U-Nets~\cite{isensee2021nnunet,ronneberger2015unet} are highly effective at capturing local and mid-range patterns, yet many failure modes in pancreas tumor segmentation are global in nature: confusing bowel enhancement for tumor, producing anatomically implausible tumor islands far from the pancreas, or missing lesions whose local intensity differs from the training data. To address these issues while preserving the multi-scale spatial structure of nnU-Net-style convolutional backbones, we incorporate a lightweight Transformer bottleneck~\cite{vaswani2017attention} at the deepest feature level. By aggregating long-range context over the global field of view, the bottleneck can encourage spatially consistent representations that generalize better when local cues shift between cohorts, complementing the locality bias that makes convolutional U-Nets stable and data-efficient.

To make these design choices meaningful, we frame evaluation explicitly around cohort transfer. We train on a recent pancreas tumor cohort~\cite{li2025pants} and test both in-cohort and out-of-cohort on an external benchmark with different data characteristics, using consistent preprocessing and reporting to avoid conflating modeling gains with pipeline differences. In addition to standard overlap metrics, we emphasize measures that reflect clinical usability under shift, including tumor sensitivity, patient-level detection behavior, and analyses stratified by tumor size and anatomical location, since cross-cohort failures often concentrate in small-lesion regimes and specific pancreatic subregions. We also examine confidence behavior and reliability, because a model that produces high-probability false positives or poorly calibrated scores can appear competitive in overlap metrics while remaining unreliable in practice.

Overall, this work argues that cohort-robust pancreas tumor segmentation can be improved by combining two complementary ideas: treating the pancreas as a probabilistic, multi-scale conditioning signal for tumor decoding, and injecting global context via a Transformer bottleneck while maintaining an nnU-Net-style convolutional backbone. The resulting approach is simple, end-to-end, and computationally practical, while directly targeting the dominant deployment pain point: generalization across cohorts, rather than optimizing only for in-distribution performance.

\section*{Results}

\subsection*{PanGuide3D architecture and comparison setting}
\begin{figure}[H]
\centering
\includegraphics[width=\linewidth]{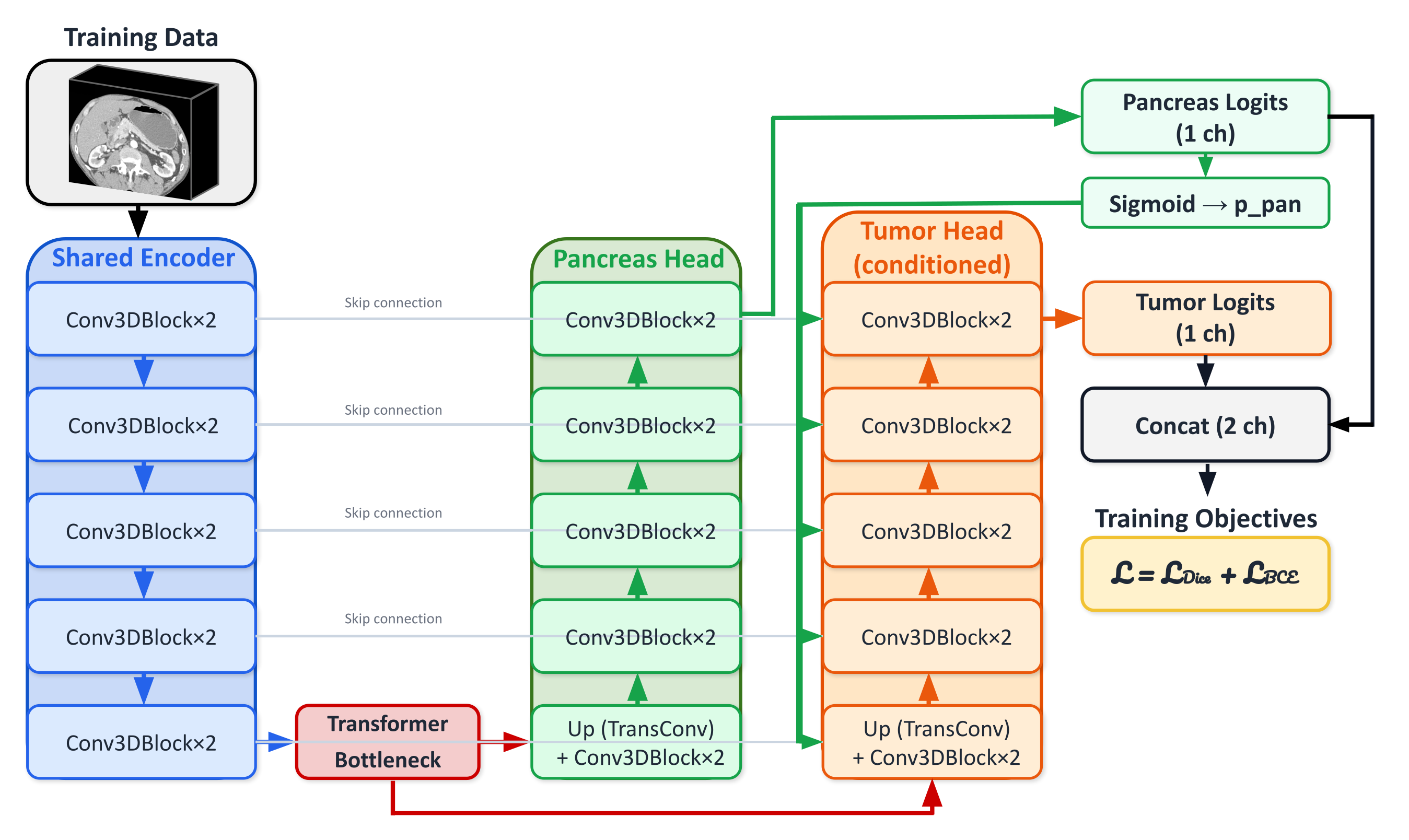}
\caption{PanGuide3D architecture. A shared 3D nnU-Net-style encoder feeds a lightweight Transformer bottleneck and two decoders: a pancreas head that predicts a probabilistic pancreas map and a tumor head conditioned on that map at multiple scales. The final output is a two-channel prediction containing pancreas and tumor logits.}
\label{fig:architecture}
\end{figure}

PanGuide3D is an nnU-Net-style 3D U-Net with a shared encoder and two task-specific decoders trained end-to-end (Figure~\ref{fig:architecture}). The encoder uses a standard multi-resolution hierarchy with repeated Conv--Norm--Activation blocks and strided downsampling. At the bottleneck, we add a lightweight Transformer module to aggregate long-range contextual information before decoding. One decoder predicts a probabilistic pancreas map, whereas the second predicts tumor logits while being conditioned on the pancreas probability map at multiple scales through soft feature modulation. The final prediction consists of concatenated pancreas and tumor logits. This design provides explicit anatomical guidance without requiring a hard cascade or a fixed region-of-interest crop.

To isolate the contributions of probabilistic organ conditioning and global contextual modeling, we evaluated two ablations. \emph{2Decoder} removes the Transformer bottleneck while retaining the shared encoder and two-decoder design with pancreas-conditioned tumor decoding, testing whether soft anatomical guidance alone explains the gains. \emph{TransBNeck} retains the Transformer bottleneck but reverts to a single-decoder formulation, testing whether global context alone can improve robustness under cohort shift without explicit organ guidance. We compare PanGuide3D and its ablations against strong baselines under matched preprocessing and training schedules: nnU-Net \cite{isensee2021nnunet}, a widely adopted self-configuring convolutional framework that remains a standard benchmark in medical segmentation; SwinUNETR \cite{hatamizadeh2022unetr}, a transformer-based alternative whose hierarchical shifted-window self-attention is designed to capture long-range context and multi-scale structure, making it a strong recent baseline for volumetric tumor segmentation; and RADFM \cite{wu2025radfm}, a recent large-scale medical foundation model trained on web-scale 2D/3D radiology data, included to test whether broad pretrained medical representations transfer effectively to this pancreas tumor segmentation setting.

\subsection*{Cohort characterization and distribution shift}
Cohort transfer for pancreas tumor segmentation is challenging not only because tumors are small and heterogeneous, but also because the underlying data distributions can differ substantially across collections. To make this distribution shift explicit, and to ensure the comparison reflects cohort differences rather than preprocessing artifacts, we harmonized PanTS \cite{li2025pants} and MSD \cite{antonelli2022msd} with the same resampling target (1.5 mm isotropic) and standardized storage format, then summarized the resulting cohorts in terms of case composition, tumor prevalence, and lesion burden.

Even after harmonization, the two cohorts present markedly different learning conditions (Table~\ref{tab:cohort_stats}). PanTS includes a large number of scans but a low tumor prevalence in the training split (684/7200, 0.0950), meaning models must learn to avoid false positives on many tumor-negative cases while still detecting subtle lesions when present; the official PanTS test split remains mixed as well (151/901, 0.1676). In contrast, MSD Task07 is entirely tumor-positive (57/57, 1.0000), so performance is dominated by how accurately tumors are delineated rather than whether the model can reject tumor-negative scans. Although both cohorts are evaluated at the same resampled resolution (1.5 mm spacing; voxel volume 3.375 mm$^3$), MSD also exhibits a substantially larger mean tumor volume (7.1168 cm$^3$) than PanTS ($\approx 2.6$ cm$^3$), indicating that cohort transfer in this setting must generalize across differences in both tumor prevalence and lesion burden.

\begin{table}[t]
\centering
\caption{Cohort statistics for PanTS and MSD after preprocessing.}
\label{tab:cohort_stats}
\small
\resizebox{\linewidth}{!}{%
\begin{tabular}{llccccc}
\toprule
Cohort & \makecell[c]{Total\\cases} & \makecell[c]{Tumor+\\cases} & \makecell[c]{Tumor\\prev. (\%)} & \makecell[c]{Spacing\\XYZ} & \makecell[c]{Voxel vol.\\(mm$^3$)} & \makecell[c]{Mean tumor vol.\\(cm$^3$)} \\
\midrule
PanTS Train & 7200 & 684 & 0.0950 & 1.5000 & 3.3750 & 2.5916 \\
PanTS Test  & 901  & 151 & 0.1676 & 1.5000 & 3.3750 & 2.6379 \\
MSD         & 57   & 57  & 1.0000 & 1.5000 & 3.3750 & 7.1168 \\
\bottomrule
\end{tabular}%
}
\end{table}

\begin{figure}[H]
\centering
\includegraphics[width=\linewidth]{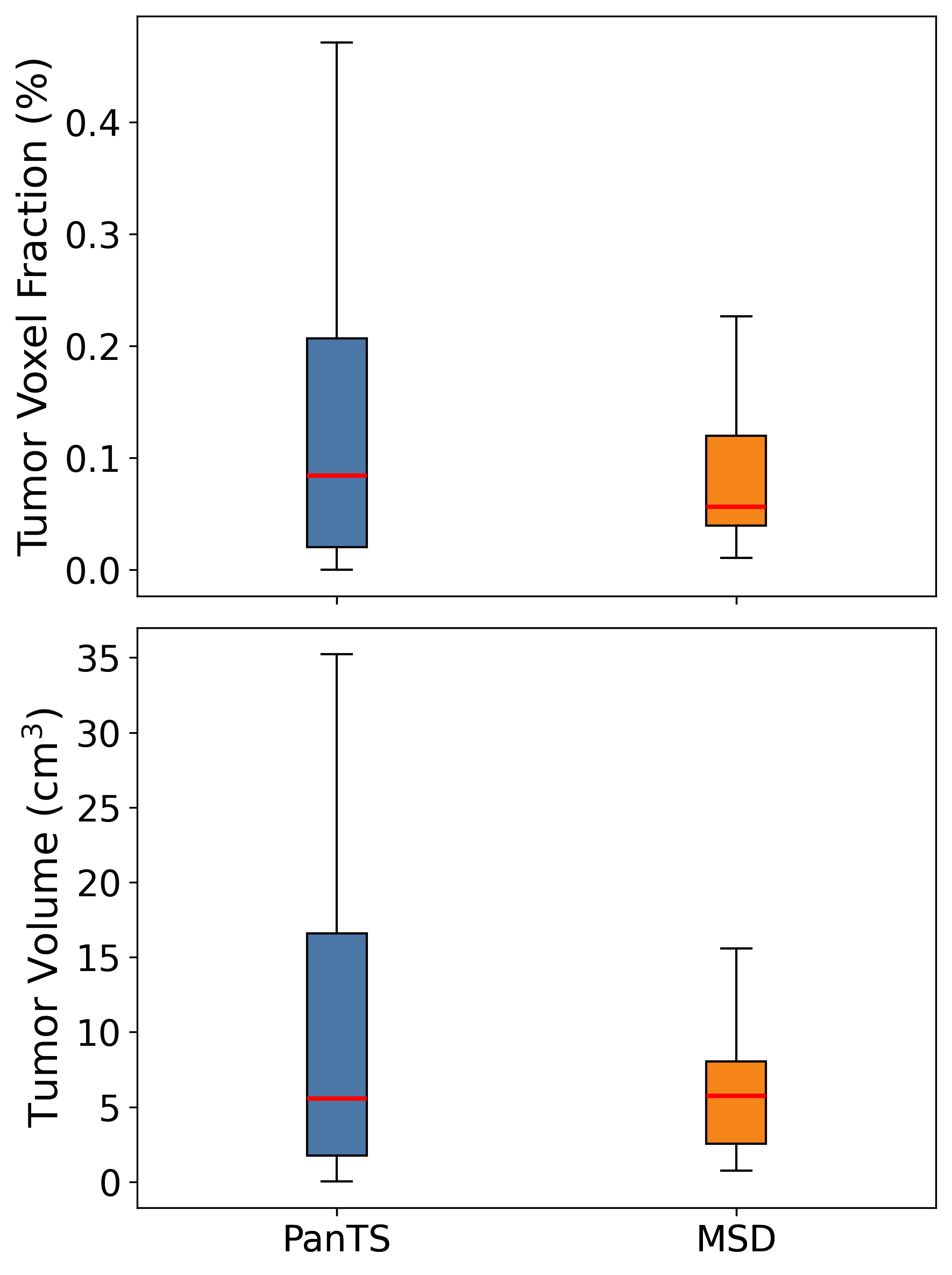}
\caption{PanTS vs. MSD tumor burden distributions. Boxplots show the cohort-wise distributions of tumor voxel fraction (top) and tumor volume (cm$^3$, bottom). PanTS exhibits a wider dynamic range with a heavier tail and many low-burden cases, while MSD concentrates around larger, more consistently sized tumors.}
\label{fig:cohort_dist}
\end{figure}

The case-level tumor burden further clarifies why cross-cohort evaluation is non-trivial. Figure~\ref{fig:cohort_dist} summarizes the per-scan distributions of tumor voxel fraction and tumor volume, highlighting that PanTS is dominated by small lesions with a heavy tail of larger tumors, while MSD is more concentrated toward higher tumor fractions and larger absolute volumes. This difference is not only a shift in summary statistics: small-lesion segmentation is particularly vulnerable to partial-volume effects, low contrast, and boundary ambiguity, where even minor localization errors can erase thin structures or fragment tiny components. Consequently, strong in-cohort performance on PanTS does not automatically translate to stable behavior under cohort shift, since the model must operate across distinct lesion-size regimes and substantially different foreground-to-background conditions.

\begin{figure}[H]
\centering
\includegraphics[width=\linewidth]{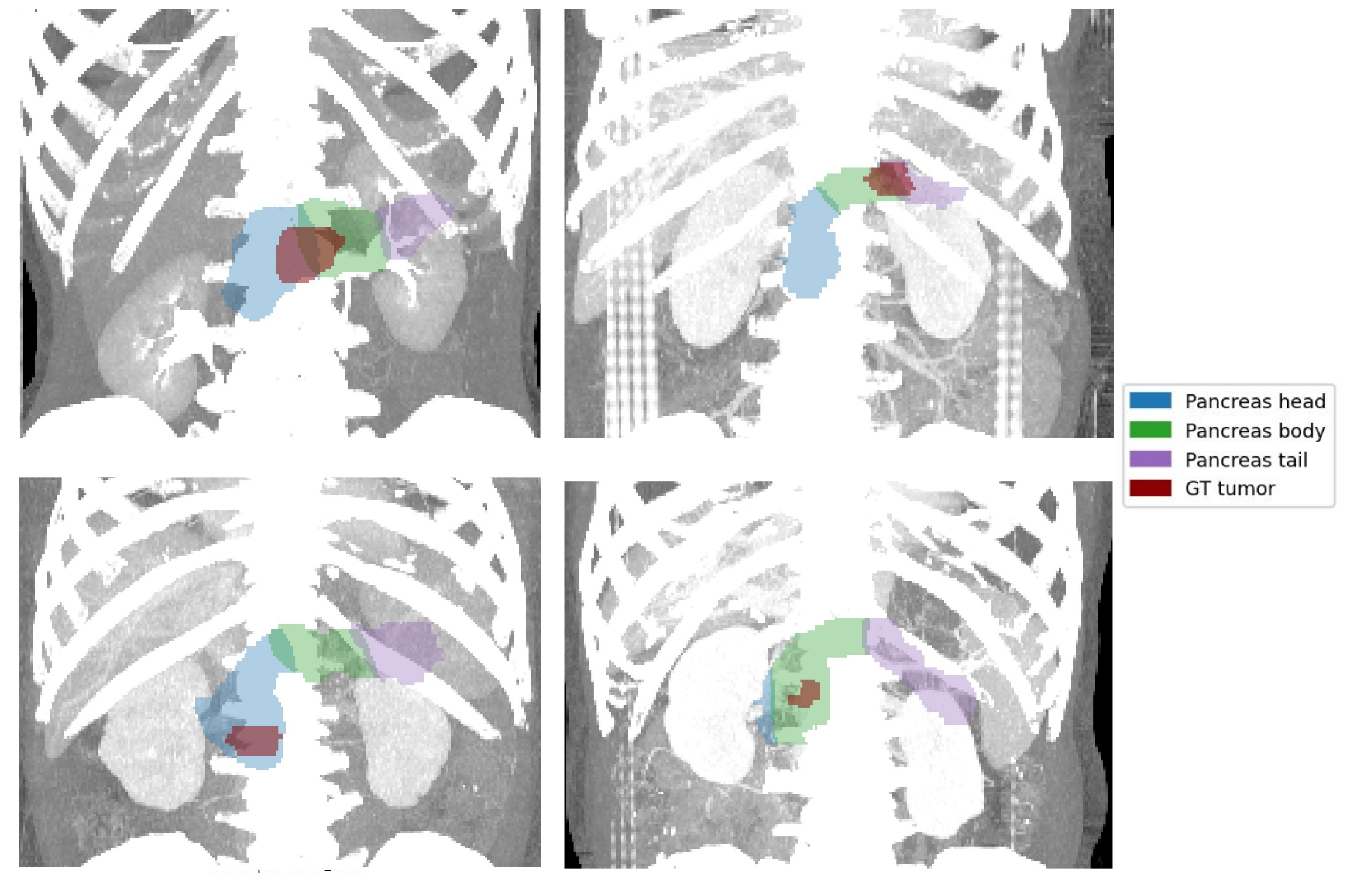}
\caption{Representative PanTS examples. CT slices illustrate the small size and heterogeneous appearance of pancreatic tumors and the close visual similarity between tumors and surrounding soft tissue. The pancreas therefore provides a useful anatomical prior for tumor localization.}
\label{fig:pants_examples}
\end{figure}

Qualitative examples help connect these statistics to what the model actually sees. The representative CT slices in Figure~\ref{fig:pants_examples} highlight the core difficulty of the task: lesions that can be visually indistinguishable from surrounding soft tissue and a pancreas whose appearance varies across patients and acquisition conditions, providing intuition for why cohort transfer can degrade performance even when spacing is matched. Importantly, PanTS also provides subregion annotations of the pancreas (head, body, and tail), enabling later stratified analyses by anatomical location in addition to standard pancreas/tumor labels.

\subsection*{Training and validation behavior}
\begin{figure}[H]
\centering
\includegraphics[width=\linewidth]{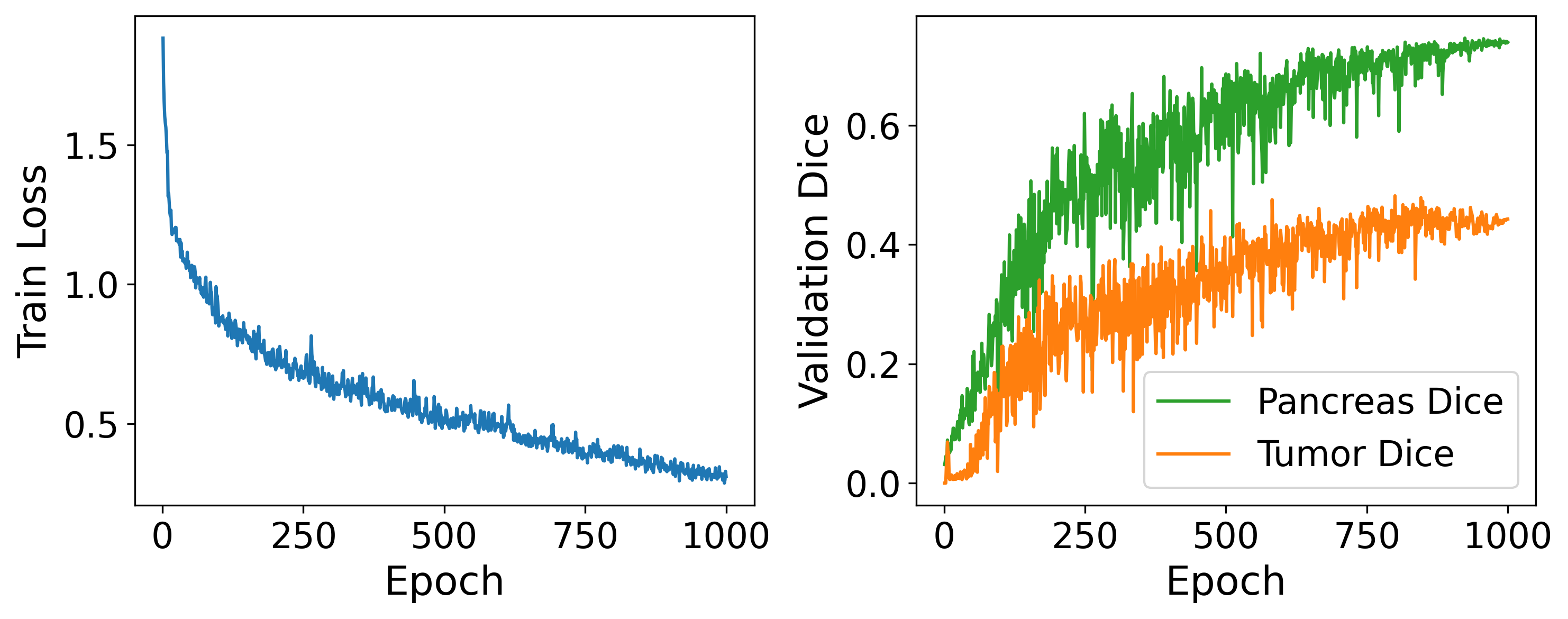}
\caption{Training dynamics for one representative fold (fold 0). Training loss decreases smoothly over epochs, whereas validation Dice improves for both tasks. Pancreas Dice rises earlier than tumor Dice, consistent with the lower difficulty of organ segmentation relative to tumor segmentation.}
\label{fig:training_dynamics}
\end{figure}

Effective cohort-robust segmentation depends on whether the model can be trained stably while learning both the easier anatomical task (pancreas) and the harder lesion task (tumor) without overfitting. Figure~\ref{fig:training_dynamics} summarizes this behavior from two complementary views: the optimization signal during training and the task-specific accuracy signal on held-out validation volumes. The training loss exhibits a smooth, monotonic decrease across epochs, indicating stable optimization under the patch-based objective and augmentation pipeline. Importantly, the curve does not show late-stage divergence or oscillatory behavior, which suggests that the learning-rate schedule and regularization (augmentation, weight decay, mixed precision with accumulation) are sufficient to prevent training instability.

The validation curves further reflect the expected difficulty gap between tasks. Pancreas Dice rises rapidly early in training and continues improving to a high plateau, consistent with the pancreas being a larger, more consistently visible structure that benefits strongly from shared low-level features. Tumor Dice improves more gradually and saturates at a lower level, reflecting the intrinsic challenges of lesion segmentation: small target size, heterogeneous appearance, and ambiguity against surrounding soft tissue. The persistent separation between the pancreas and tumor trajectories is therefore not a failure mode, but an empirical signature of task imbalance and the harder decision boundary for tumors.

\subsection*{Cohort transfer performance and component analysis}
\begin{figure}[H]
\centering
\includegraphics[width=\linewidth]{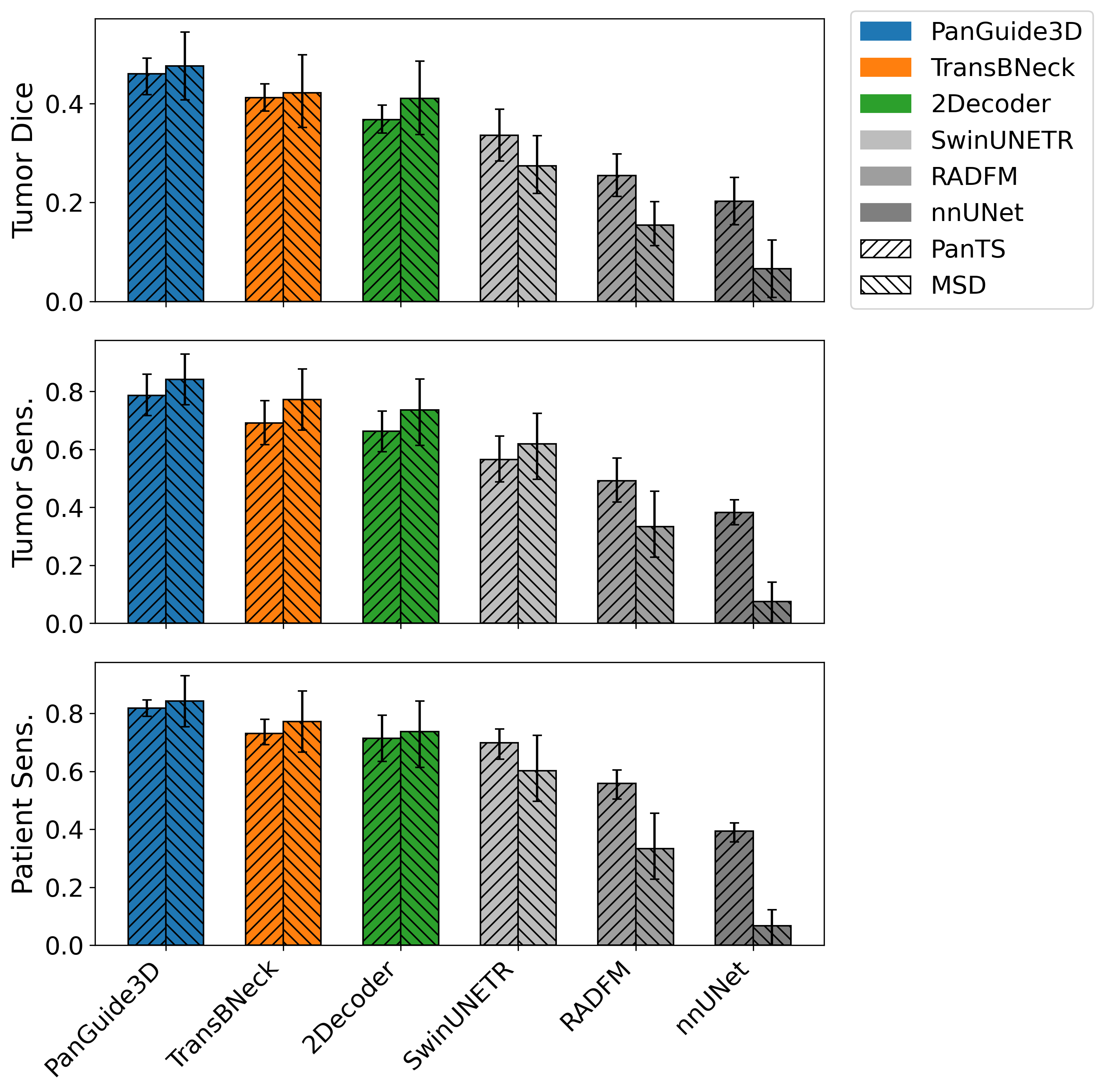}
\caption{Tumor segmentation performance on PanTS (in-cohort) and MSD (out-of-cohort). Bars show mean $\pm$ standard deviation across folds for tumor Dice, tumor sensitivity, and patient sensitivity. PanGuide3D shows the highest mean performance among the evaluated models across these metrics.}
\label{fig:main_performance}
\end{figure}

We next evaluated whether PanGuide3D improves pancreas tumor segmentation accuracy while remaining robust under cohort shift, using a cohort-transfer setting that reflects deployment reality: training on PanTS and testing both in-cohort (PanTS) and out-of-cohort (MSD). Figure~\ref{fig:main_performance} summarizes cohort-transfer performance with three complementary tumor-focused metrics: tumor Dice (overlap quality), tumor sensitivity (voxel-level recall), and patient sensitivity (case-level detection among tumor-positive scans). Error bars indicate standard deviation across five folds, and the side-by-side layout makes the generalization gap and relative ranking of architectures immediately comparable.

A clear pattern across baselines is that segmentation quality is not stable under cohort shift: several methods that are competitive on PanTS deteriorate substantially on MSD, despite matched preprocessing. PanGuide3D shows the highest mean performance across the three panels in this comparison, with strong tumor Dice and sensitivity on both cohorts, suggesting that the gains are not limited to a single operating point but extend to both overlap quality and detection behavior.

The sensitivity panels complement Dice by capturing additional aspects of detection performance. Tumor sensitivity measures whether tumor voxels are recovered, while patient sensitivity captures whether a tumor is detected at all in a positive case, an especially relevant criterion for deployment where missing a lesion is costly. The gap between these metrics for baselines (SwinUNETR and RADFM) suggests brittle behavior: models may recover portions of tumor when they lock on, yet still fail to consistently trigger on challenging cases under shift. PanGuide3D narrows this gap, improving both voxel-level recall and patient-level detection, consistent with fewer shift-induced miss failures.

The ablation experiments help interpret these gains. The \emph{2Decoder} variant, which retains pancreas-guided tumor decoding but removes the Transformer bottleneck, already improves tumor Dice and sensitivity relative to single-head baselines, supporting the role of multi-scale pancreas conditioning as a soft anatomical constraint. The \emph{TransBNeck} variant, which adds global context without explicit pancreas guidance, also improves robustness relative to the purely convolutional baseline, suggesting that long-range contextual cues help when appearance statistics shift across cohorts. The full PanGuide3D model performs better than either ablation, consistent with anatomical guidance and global context contributing complementary benefits.

Relative to other strong baselines, the trends are consistent with their architectural priors. SwinUNETR benefits from stronger global modeling than nnU-Net but can still underperform in tumor metrics under shift, while RADFM may not translate the representation advantages into reliable tumor delineation in this setting. Overall, Figure~\ref{fig:main_performance} supports the central claim that end-to-end probabilistic anatomical conditioning, augmented by a lightweight Transformer bottleneck, yields a more transferable tumor segmentation model than either component alone.

Table~\ref{tab:main_results} provides detailed numerical results (mean $\pm$ standard deviation) and complements Figure~\ref{fig:main_performance} by adding pancreas Dice and false-positive tumor volume. It reinforces two key points: first, plain nnU-Net can fail catastrophically in this cohort-transfer setting, producing near-zero tumor Dice and large false-positive volumes; second, stronger architectural priors or training variants (for example, TransBNeck and 2Decoder) substantially improve both overlap and sensitivity while reducing false positives. The table also highlights trade-offs among competing methods; for example, SwinUNETR achieves strong pancreas Dice yet remain weak on tumor delineation, underscoring why pancreas-aware tumor decoding is valuable when the goal is robust tumor segmentation rather than organ segmentation alone.

\begin{table}[H]
\centering
\caption{Main quantitative results across cohorts and models (mean $\pm$ standard deviation across folds).}
\label{tab:main_results}
\small
\resizebox{\linewidth}{!}{%
\begin{tabular}{llccccc}
\toprule
Model & Cohort & T Dice $\uparrow$ & P Dice $\uparrow$ & FP volume (cm$^3$) $\downarrow$ & Tumor sens. $\uparrow$ & Patient sens. $\uparrow$ \\
\midrule
nnUNet     & MSD   & 0.066$\pm$0.057 & 0.145$\pm$0.076 & 19.646$\pm$5.965 & 0.075$\pm$0.065 & 0.068$\pm$0.057 \\
RADFM      & MSD   & 0.154$\pm$0.048 & 0.489$\pm$0.033 & 8.666$\pm$1.387  & 0.333$\pm$0.123 & 0.333$\pm$0.123 \\
SwinUNETR  & MSD   & 0.274$\pm$0.061 & 0.761$\pm$0.024 & 8.550$\pm$1.542  & 0.619$\pm$0.105 & 0.602$\pm$0.102 \\
2Decoder   & MSD   & 0.410$\pm$0.075 & $\mathbf{0.772}\pm0.022$ & 4.781$\pm$1.595  & 0.737$\pm$0.106 & 0.737$\pm$0.106 \\
TransBNeck & MSD   & 0.421$\pm$0.077 & 0.749$\pm$0.026 & 4.291$\pm$1.515  & 0.772$\pm$0.106 & 0.772$\pm$0.106 \\
\textbf{PanGuide3D} & MSD   & $\mathbf{0.475}\pm0.069$ & 0.753$\pm$0.033 & $\mathbf{2.004}\pm1.765$  & $\mathbf{0.842}\pm0.088$ & $\mathbf{0.842}\pm0.088$ \\
\midrule
nnUNet     & PanTS & 0.202$\pm$0.054 & 0.305$\pm$0.090 & 16.970$\pm$6.107 & 0.383$\pm$0.043 & 0.394$\pm$0.049 \\
RADFM      & PanTS & 0.255$\pm$0.042 & 0.559$\pm$0.010 & 10.835$\pm$0.338 & 0.491$\pm$0.079 & 0.559$\pm$0.011 \\
SwinUNETR  & PanTS & 0.336$\pm$0.053 & $\mathbf{0.827}\pm0.009$ & 9.883$\pm$0.452  & 0.565$\pm$0.082 & 0.699$\pm$0.057 \\
2Decoder   & PanTS & 0.367$\pm$0.029 & 0.708$\pm$0.016 & 7.828$\pm$0.820  & 0.663$\pm$0.070 & 0.714$\pm$0.089 \\
TransBNeck & PanTS & 0.412$\pm$0.027 & 0.683$\pm$0.014 & 6.773$\pm$0.699  & 0.692$\pm$0.077 & 0.731$\pm$0.048 \\
\textbf{PanGuide3D} & PanTS & $\mathbf{0.460}\pm0.041$ & 0.804$\pm$0.013 & $\mathbf{3.070}\pm1.554$  & $\mathbf{0.787}\pm0.073$ & $\mathbf{0.819}\pm0.039$ \\
\bottomrule
\end{tabular}%
}
\end{table}

\subsection*{Tumor size sensitivity analysis}
To understand when segmentation models succeed or fail, and whether performance degrades disproportionately on clinically challenging small lesions, we analyzed how tumor Dice varies with tumor volume. Figure~\ref{fig:volume_scatter} plots per-case tumor Dice against tumor volume (cm$^3$) for PanGuide3D and nnU-Net on both the in-cohort PanTS test set and the out-of-cohort MSD cohort. Each point corresponds to one scan, allowing inspection of size-dependent behavior beyond aggregate averages. PanGuide3D shows a broad band of moderate-to-high Dice scores across a wide range of tumor volumes, while nnU-Net points cluster near very low Dice for many small tumors and only occasionally reach high Dice at larger volumes.

\begin{figure}[H]
\centering
\includegraphics[width=\linewidth]{figures/Figure_6.png}
\caption{Tumor Dice vs. tumor volume (cm$^3$) for one representative fold (fold 0). Small lesions are substantially harder, exhibiting wide Dice variability and frequent near-zero failures. PanGuide3D maintains higher Dice across a broader range of tumor sizes.}
\label{fig:volume_scatter}
\end{figure}

This size-stratified view indicates that tumor volume is a strong driver of segmentation difficulty, especially under cohort transfer where small lesions are more easily missed and boundary errors dominate. Relative to nnU-Net, PanGuide3D sustains noticeably higher Dice in the low-volume regime, suggesting better detection and delineation when tumors occupy only a small fraction of voxels and contrast is limited. On MSD, nnU-Net exhibits a marked collapse toward near-zero Dice for many cases, whereas PanGuide3D retains a consistent subset of higher Dice outcomes across typical tumor sizes, aligning with improved robustness under distribution shift and fewer catastrophic failures on hard, small-lesion cases.

\subsection*{Stratified performance by tumor size and pancreatic subregion}
\begin{figure}[H]
\centering
\includegraphics[width=\linewidth]{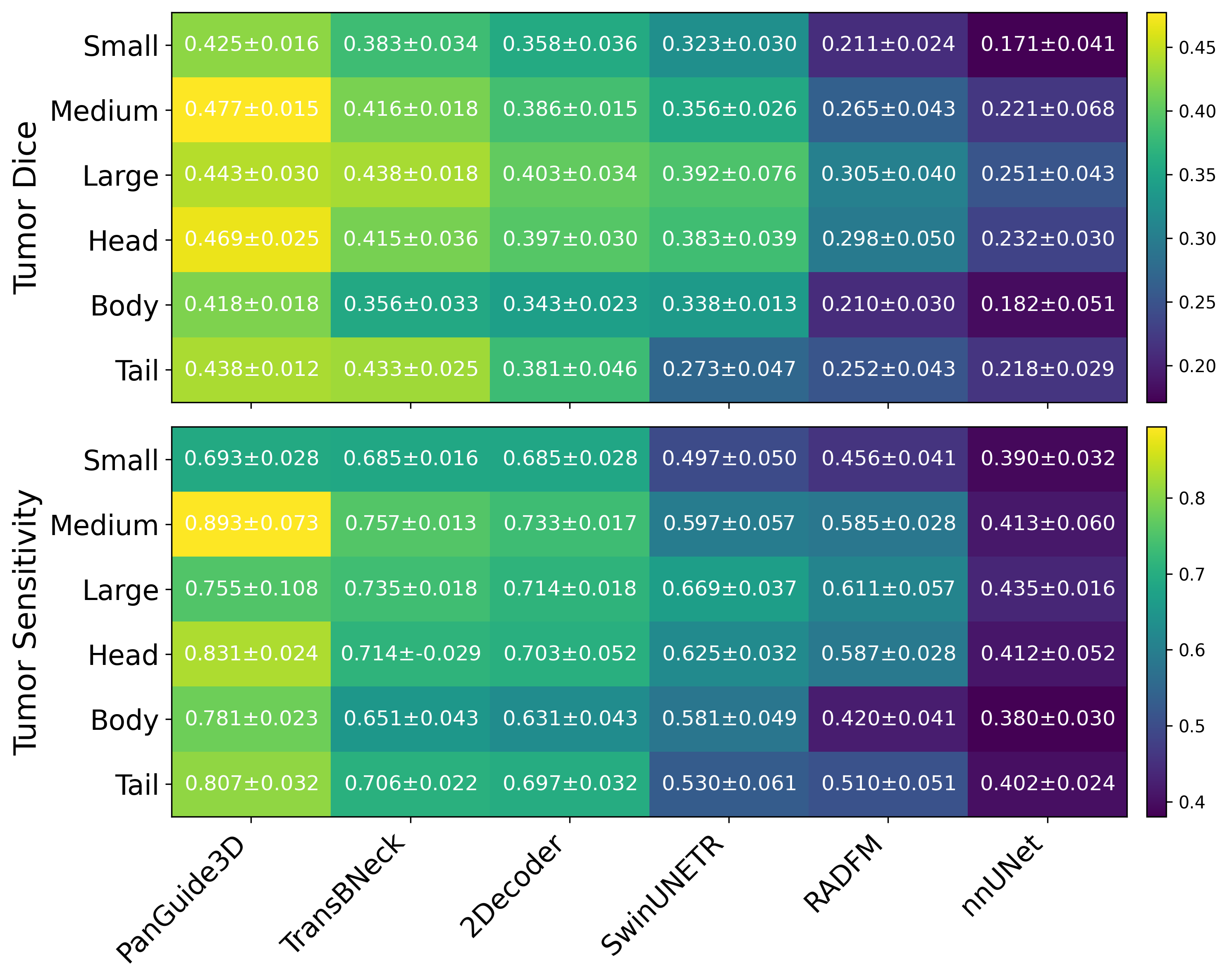}
\caption{Stratified subgroup performance on PanTS (size and location). Heatmaps report tumor Dice and tumor sensitivity with standard deviation across folds. PanGuide3D is strongest in challenging strata such as small tumors and body/tail regions.}
\label{fig:heatmap}
\end{figure}

To probe where cohort-robust improvements arise beyond a single average Dice, we report stratified performance across clinically meaningful strata. Figure~\ref{fig:heatmap} summarizes two heatmaps: the top shows tumor Dice and the bottom shows tumor sensitivity, each reported as mean $\pm$ standard deviation for every model over five-fold cross-validation. Rows stratify cases by tumor size (small, medium, and large) and by pancreatic subregion (head, body, and tail), while columns compare PanGuide3D against representative baselines.

Several patterns emerge. First, performance gaps are largest in the most challenging regimes: small tumors and body, where sensitivity is typically the bottleneck. PanGuide3D is consistently strongest or tied among the best across most strata, with particularly clear advantages in tumor Dice and sensitivity for medium/large tumors and for head/tail cases, indicating that probabilistic organ-guided decoding helps the model remain anatomically focused while still capturing lesion boundaries. Second, the baselines tend to degrade unevenly across strata: nnU-Net shows extremely low Dice and sensitivity for small tumors and markedly weaker results in body cases, suggesting frequent misses and instability under limited lesion signal. Similarly, RADFM and SwinUNETR appear more brittle in the harder strata. Overall, the stratified view supports the conclusion that PanGuide3D’s gains are not driven by a single easy subset. Instead, the model improves robustness across tumor size and pancreatic subregions, which is critical for reliable deployment under distribution shift.

\subsection*{Qualitative comparison of organ-guided tumor segmentation}
\begin{figure}[H]
\centering
\includegraphics[width=\linewidth]{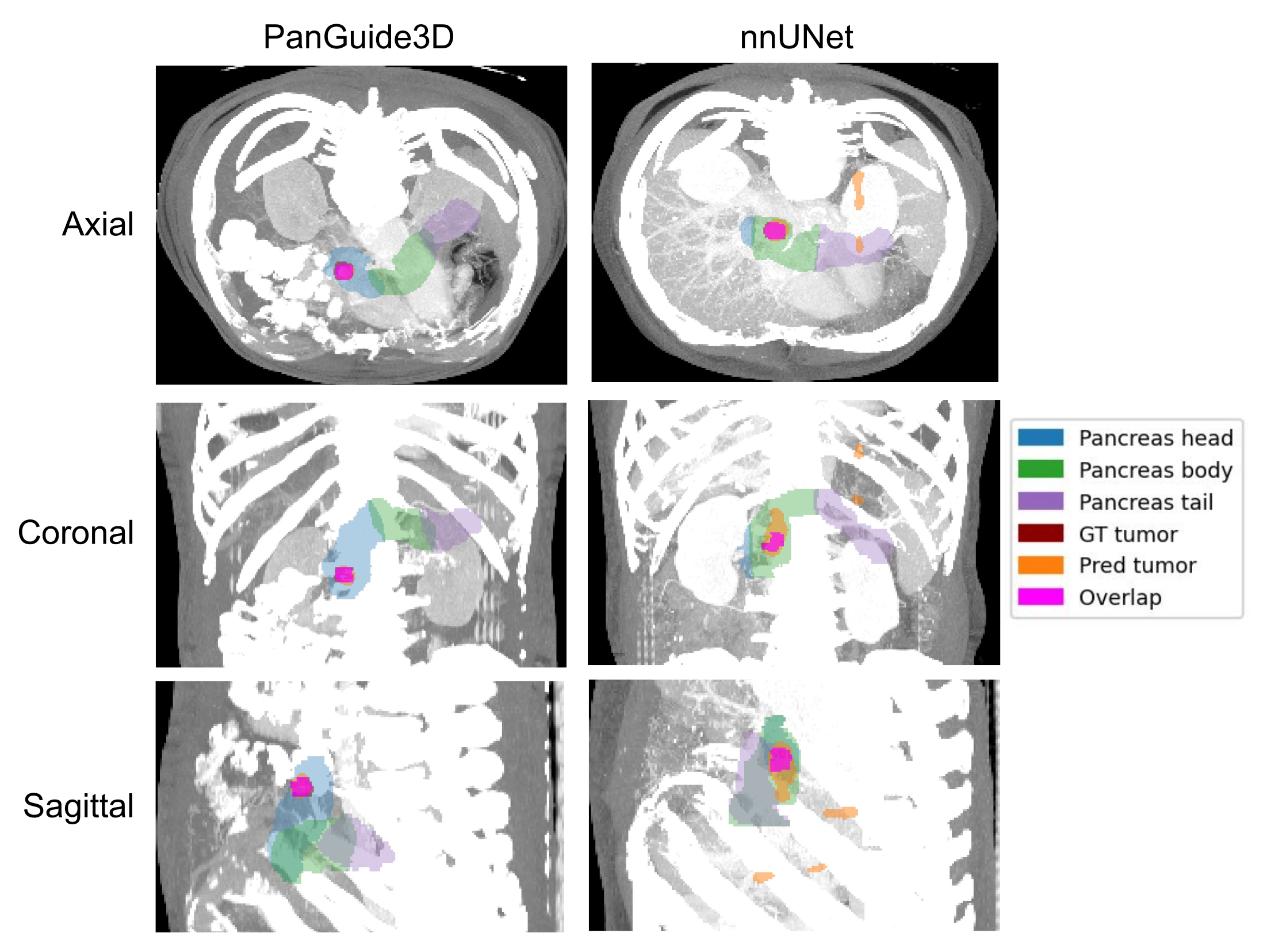}
\caption{Qualitative comparison. PanGuide3D produces more anatomically plausible tumor predictions with fewer off-site false positives, while single-head nnU-Net can generate spurious tumor regions outside the pancreas under cohort shift.}
\label{fig:qualitative}
\end{figure}

To complement the quantitative results, Figure~\ref{fig:qualitative} provides a side-by-side qualitative comparison between PanGuide3D and the nnU-Net baseline, visualized across axial, coronal, and sagittal views. The pancreas is shown with subregion labels (head/body/tail), while the tumor annotations include ground truth, prediction, and their overlap. PanGuide3D produces a compact tumor prediction that aligns with the ground-truth lesion and remains anatomically consistent with the pancreas mask across all three planes. In contrast, nnU-Net generates additional predicted tumor regions that extend outside the pancreas anatomy, indicating off-site false positives that violate the expected organ prior and are consistent with the elevated false-positive behavior observed under cohort shift.

\subsection*{Confidence calibration under cohort shift}
\begin{figure}[H]
\centering
\includegraphics[width=\linewidth]{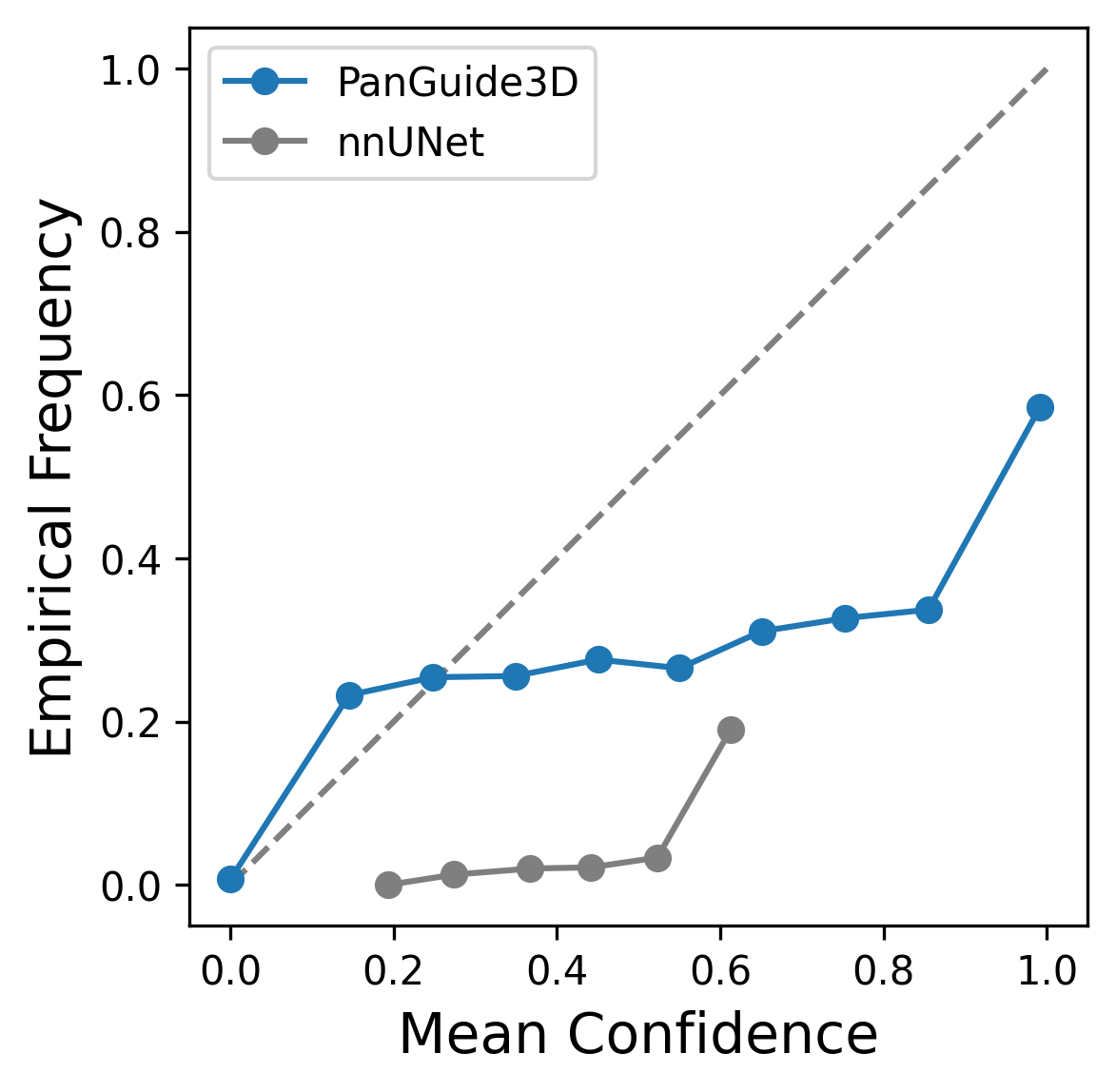}
\caption{Reliability diagram for tumor-probability calibration (PanTS$\rightarrow$MSD) for one representative fold (fold 0). Empirical tumor frequency is plotted against mean predicted confidence after binning voxel-wise tumor probabilities; the dashed diagonal denotes perfect calibration.}
\label{fig:calibration}
\end{figure}

To assess whether a model’s predicted probabilities can be trusted as confidence, especially under cohort shift, we constructed the reliability diagram in Figure~\ref{fig:calibration}, which compares predicted confidence to empirical correctness. Concretely, we take voxel-wise tumor probabilities from each model, partition voxels into confidence bins, and for each bin compute the mean confidence and the empirical frequency of tumor voxels. Plotting empirical frequency versus mean confidence yields a calibration curve, where the dashed diagonal indicates perfect calibration. In this plot, PanGuide3D tracks the diagonal more closely than nnU-Net across most confidence bins, indicating more reliable probability estimates under cohort shift, but it remains over-confident at the highest-confidence bin(s) where predicted confidence exceeds the observed empirical frequency. In contrast, nnU-Net remains near zero empirical frequency across a wide range of moderate confidence values and only rises at higher confidence, suggesting miscalibration where intermediate confidence scores do not correspond to consistent correctness. This calibration improvement implies that PanGuide3D provides more meaningful confidence signals for practical use, such as selecting probability thresholds, triaging uncertain regions or cases, or supporting uncertainty-aware review.

\section*{Discussion}
In this work, we presented PanGuide3D, an end-to-end 3D nnU-Net-style framework that improves cohort robustness by combining a shared encoder, a pancreas decoder, and a tumor decoder explicitly conditioned on multi-scale pancreas probabilities, together with a lightweight Transformer bottleneck to strengthen long-range contextual reasoning. Across in-cohort PanTS evaluation and out-of-cohort transfer to MSD Task07, PanGuide3D consistently delivers stronger tumor overlap and detection behavior than strong baselines and modern architectures, while reducing anatomically implausible false positives and maintaining competitive pancreas segmentation. These results suggest that probabilistic anatomical conditioning is a practical mechanism to stabilize tumor predictions under distribution shift, with potential applications in clinical workflows such as preoperative planning, longitudinal tumor monitoring, and decision support where reliable segmentation can reduce manual burden and improve reproducibility.

Despite these gains, several limitations remain. First, the model outputs should be treated as decision support rather than definitive clinical conclusions. Performance can vary with image quality, scanner protocols, and rare tumor presentations, and model errors may still occur even when confidence appears high. Second, while cohort robustness improves, tumor Dice and boundary accuracy can still be further strengthened, particularly for very small lesions where partial-volume effects and annotation ambiguity dominate the error budget. Third, the calibration analysis indicates that PanGuide3D can be over-confident at high confidence bins, which means probability scores may not always correspond to the true reliability needed for automated thresholding or downstream triage without additional safeguards. Finally, the evaluation is limited to the available cohorts and labels. Broader multi-institutional testing would be needed to fully characterize real-world generalization.

Future work can improve both accuracy and trustworthiness in several directions. On the modeling side, stronger but still efficient context modules, richer pancreas-to-tumor conditioning, and loss designs that better emphasize small-lesion delineation can be further explored. On the training side, expanding multi-cohort training and adding domain generalization strategies such as intensity harmonization, style perturbations, or domain-adversarial regularization may further reduce transfer gaps. On the deployment side, improving uncertainty estimation and calibration can make confidence signals more reliable for clinical review. Together, these advances would move PanGuide3D toward a more accurate, robust, and clinically actionable pancreas tumor segmentation system.

\section*{Methods}
\subsection*{Computational resources and software}
All experiments were run on a Linux system equipped with eight NVIDIA A100 GPUs (80 GB). The training and inference pipeline was implemented in PyTorch, with selected components and utilities from MONAI. Mixed-precision execution was enabled where applicable to improve throughput and reduce memory usage, and a single shared evaluation pipeline was used across all models.

We relied on the following core software packages throughout the workflow: NumPy, SciPy, and Pandas for data handling; PyTorch and MONAI for model implementation and training; scikit-image for image utilities; SimpleITK for medical image I/O and metadata handling; and Matplotlib for plotting. Optional 3D visualization and rendering were performed with PyVista.

\subsection*{Data collection and preprocessing}
We conducted experiments on two public pancreas tumor segmentation cohorts with different acquisition characteristics and tumor burden: PanTS~\cite{li2025pants} and MSD Task07 Pancreas~\cite{antonelli2022msd}. For PanTS, we started from the released raw contrast-enhanced CT volumes and corresponding annotations, resampled both images and labels to a uniform isotropic spacing of 1.5 mm, and saved each case as a standardized \texttt{.npz} file containing the image volume, multi-channel label volume, and voxel spacing metadata. Using the PanTS training split (9,000 cases), we created an internal development split by randomly partitioning cases into 80\% training and 20\% validation, resulting in 7,200 training volumes and 1,800 validation volumes, while keeping the official PanTS test set (901 cases) fully held out for final evaluation. For MSD Task07 Pancreas, we similarly began from the raw CT scans and segmentation masks provided by the Medical Segmentation Decathlon and converted them into the same \texttt{.npz} representation and preprocessing pipeline as PanTS, ensuring consistent data formatting and spatial resolution across cohorts to enable fair cohort-transfer evaluation and minimize confounding differences due to preprocessing.

\subsection*{Compared methods and ablations}
The PanGuide3D architecture is summarized in the first subsection of the Results and in Figure~\ref{fig:architecture}. Here, we focus on the compared methods and ablation design used for evaluation. To isolate the contributions of probabilistic organ conditioning and global context modeling, we evaluated two ablations. \emph{2Decoder} removes the Transformer bottleneck while retaining the shared encoder and pancreas-conditioned tumor decoding, thereby testing whether anatomical guidance alone explains the gains. \emph{TransBNeck} keeps the Transformer bottleneck but reverts to a single-decoder formulation without a separate pancreas head or explicit conditioning, thereby testing whether global context alone improves robustness under cohort shift.

We compared PanGuide3D against strong baselines and representative 3D segmentation models under matched preprocessing and training schedules. nnU-Net~\cite{isensee2021nnunet} served as the primary convolutional baseline. SwinUNETR~\cite{hatamizadeh2022unetr} provided a transformer-based alternative with stronger global modeling, and RADFM~\cite{wu2025radfm} was included as a foundation-model-based baseline. The two-stage U-Net method~\cite{ghorpade2024twostage} does not release training code, so it could not be reproduced under our matched experimental setting.

\subsection*{Model training and hyperparameter optimization}
Models were trained with a joint segmentation objective on pancreas and tumor channels using a Dice plus binary cross-entropy (BCE) loss, applied per channel and summed, with focal and Tversky variants explored in pilot trials. Let $p_{c,i}$ denote the predicted probability for voxel $i$ in channel $c$ and let $y_{c,i}$ denote the corresponding binary ground-truth label. We define the soft Dice score for channel $c$ as
\begin{equation}
\mathrm{Dice}(p_c,y_c)=\frac{2\sum_i p_{c,i}y_{c,i}+\varepsilon}{\sum_i p_{c,i}+\sum_i y_{c,i}+\varepsilon},
\end{equation}
with Dice loss
\begin{equation}
\mathcal{L}_{\mathrm{Dice},c}=1-\mathrm{Dice}(p_c,y_c).
\end{equation}
The BCE loss for channel $c$ is
\begin{equation}
\mathcal{L}_{\mathrm{BCE},c}=-\frac{1}{N}\sum_i\left[y_{c,i}\log p_{c,i}+(1-y_{c,i})\log(1-p_{c,i})\right],
\end{equation}
where $N$ is the number of voxels in the patch. The overall multitask objective sums losses across channels,
\begin{equation}
\mathcal{L}=\sum_{c\in\{\mathrm{pancreas},\mathrm{tumor}\}}\left(\mathcal{L}_{\mathrm{Dice},c}+\lambda\,\mathcal{L}_{\mathrm{BCE},c}\right),
\end{equation}
where $\lambda$ controls the BCE weight.

Training is performed on 3D patches cropped from each CT volume (default size $160\times160\times96$), which improves memory efficiency and increases the diversity of local contexts seen per epoch. To address extreme class imbalance, we bias sampling toward tumor-centered patches via a configurable tumor sampling probability and further balance tumor-positive and tumor-negative cases with a weighted sampler. We apply standard 3D medical imaging augmentations, including Gaussian noise, blur, brightness/contrast jitter, gamma transforms, and low-resolution simulation, to improve robustness. In practice, augmentation is essential for stable tumor learning. All methods use mixed-precision training to accelerate throughput, and we enable gradient accumulation when GPU memory limits batch size.

Optimization uses AdamW with weight decay by default, while nnU-Net follows its recommended configuration using stochastic gradient descent with momentum and Nesterov acceleration for a faithful baseline. Unless otherwise specified, we apply polynomial learning-rate decay over update steps,
\begin{equation}
\eta_t = \eta_0\left(1-\frac{t}{T}\right)^{0.9},
\end{equation}
where $\eta_0$ is the initial learning rate, $t$ is the current update step, and $T$ is the total number of update steps. We log step-level training losses and perform epoch-level validation using sliding-window inference to match test-time behavior. Model selection is based on validation tumor Dice, and we keep preprocessing, augmentation, and training schedules consistent across PanGuide3D and all baselines to ensure fair comparisons. Hyperparameter exploration focuses on learning rate, learning-rate strategy (fixed versus polynomial decay), optimizer choice, weight decay, and augmentation settings, as these factors most strongly influence convergence and cross-cohort generalization.

\subsection*{Evaluation}
We evaluated PanGuide3D and all baselines using five-fold cross-validation on the PanTS training set and a consistent post-processing and metric pipeline across cohorts. For each fold, we trained on the fold's training split and selected checkpoints using the corresponding validation split; we then evaluated the selected model on the official PanTS test set and the MSD cohort. For each case, we loaded the preprocessed \texttt{.npz} volume and the model's saved prediction probability maps from disk, then converted probabilities to binary masks using a chosen threshold $\tau$. We report voxel-wise segmentation quality for both tumor and pancreas, together with error and detection-oriented metrics that better reflect clinical usability under cohort shift.

Tumor sensitivity (recall) is defined as
\begin{equation}
\mathrm{Tumor\ Sensitivity}=\frac{\mathrm{TP}}{\mathrm{TP}+\mathrm{FN}},
\end{equation}
where TP and FN denote true-positive and false-negative tumor voxels, respectively. Patient sensitivity is defined as the fraction of tumor-positive patients for which the model detects at least one tumor voxel:
\begin{equation}
\mathrm{Patient\ Sensitivity}=\frac{1}{N_{+}}\sum_{n=1}^{N_{+}}\mathbb{I}\left(\sum_i \hat{y}_{n,i}>0\right),
\end{equation}
where $N_{+}$ is the number of tumor-positive cases, $\hat{y}_{n,i}$ denotes the binarized tumor prediction for voxel $i$ in patient $n$, and $\mathbb{I}(\cdot)$ is the indicator function. In addition to tumor Dice and sensitivity, we report pancreas Dice, false-positive tumor volume, size-stratified analyses, location-stratified analyses, and calibration behavior.

\section*{Data availability}
The PanTS and MSD Task07 Pancreas datasets analyzed in this study are publicly available from their respective organizers and repositories, subject to the terms and conditions of those datasets. Processed derivatives used for analysis can be made available from the corresponding author upon reasonable request and in accordance with dataset licensing restrictions.

\section*{Code availability}
Code used for preprocessing, model training, evaluation, and figure generation is available from the corresponding author upon reasonable request. A public release may be considered after manuscript finalization, subject to institutional and project constraints.

\section*{Acknowledgements}
The authors acknowledge the computational resources that supported model training and evaluation.

\section*{Author contributions}
S.M. contributed to conceptualization, literature review, methodology development, computational experiments, data analysis, figure preparation, and preparation of the initial manuscript draft. X.M. contributed to study supervision, methodology development, data analysis, and manuscript revision. All authors reviewed and approved the final manuscript.

\section*{Competing interests}
The authors declare no competing interests.

\bibliographystyle{unsrt}
\bibliography{sn-bibliography}

\end{document}